\documentclass{ceurart}

\usepackage{graphicx}
\usepackage{booktabs}
\usepackage{amsmath}
\usepackage{url}
\usepackage{xcolor}
\usepackage{placeins}
\usepackage{microtype}
\usepackage{enumitem}

\usepackage[numbers]{natbib}

\begin{document}

\copyrightyear{2026}
\copyrightclause{Copyright for this paper by its authors.
  Use permitted under Creative Commons License Attribution 4.0
  International (CC BY 4.0).}

\conference{CLEF 2026 Working Notes, 21 -- 24 September 2026, Jena, Germany}

\title{
KhyontekAI at CheckThat! 2026: Multilingual Dense Retrieval and Cross-Encoder Reranking for Scientific Web Claims
}

\author[1]{Bhargab Jyoti Bhuyan}[%
  email=bjbcr7@gmail.com,
  orcid=0009-0008-2204-4086
]

\author[2]{Pritam Deka}[%
  email=pritam@khyontekai.com,
  orcid=0000-0002-8655-4762
]

\address[1]{Computer Science and Engineering Department, School of Engineering, Tezpur University, Assam, India}

\address[2]{KhyontekAI, Guwahati, Assam, India}

\begin{abstract}
Scientific claims shared through social media platforms frequently reference research findings without explicitly linking to the corresponding scientific publications. Automatically retrieving the correct scientific source for such claims is a challenging multilingual retrieval problem involving semantic matching, paraphrasing, noisy social media text, and cross-domain language variation. In this work, we present a multilingual two-stage retrieval framework for CLEF 2026 CheckThat! Task 1 on Source Retrieval for Scientific Web Claims. Our system combines fine-tuned dense retrieval models with cross-encoder reranking for improving multilingual scientific document retrieval. We evaluate multiple transformer-based dense retrieval architectures including multilingual-E5, BGE, and Contriever under different metadata configurations. We further investigate multilingual joint training, retrieval depth analysis, metadata enrichment, and language balancing strategies for improving retrieval effectiveness. Experimental results demonstrate that metadata enrichment consistently improves retrieval effectiveness, while fine-tuned cross-encoder reranking provides additional gains in retrieval precision. Our final multilingual system achieves MRR@5 scores of 0.6222 for English, 0.5112 for German, and 0.6254 for French, with an overall average MRR@5 score of 0.5862.
\end{abstract}


\maketitle

\section{Introduction}

The rapid growth of social media platforms has significantly changed the way scientific information is disseminated and consumed online \cite{scientificsurvey,evidencefactchecking}. Scientific findings are now frequently summarized, discussed, and reshared through short-form posts, blogs, forums, and news articles. However, these claims are often presented without direct references to the original scientific publications, making verification difficult for both users and automated fact-checking systems. The increasing spread of scientific misinformation and misleading interpretations further highlights the importance of reliable evidence retrieval systems capable of linking claims to trustworthy scientific sources.

Recent advances in automated fact-checking have emphasized the importance of evidence retrieval as a critical first stage in verification pipelines \cite{scientificsurvey,evidencefactchecking}. Prior studies in scientific fact-checking and evidence-based verification have shown that retrieval quality strongly influences downstream verification performance. In multilingual environments, this problem becomes substantially more challenging due to linguistic variability, paraphrasing, domain-specific terminology, noisy user-generated text, and semantic differences across languages.

The CLEF 2026 CheckThat! Lab Task 1 focuses on the problem of source retrieval for scientific web claims \cite{checkthat2026task1}. Given a multilingual scientific claim extracted from online content, the task requires retrieving the corresponding scientific publication from a large collection of candidate documents. The retrieval system must therefore perform robust semantic matching between noisy claims and structured scientific metadata while generalizing across multiple languages including English, German, and French.

Traditional sparse retrieval methods such as BM25 \cite{bm25} rely heavily on lexical overlap and exact term matching, which often limits their effectiveness for scientific claim retrieval where paraphrasing and semantic reformulation are common. Dense retrieval approaches based on transformer embeddings have recently demonstrated substantial improvements in semantic retrieval tasks by encoding queries and documents into dense vector representations. Models such as Sentence-BERT \cite{sbert}, Contriever \cite{contriever}, E5 \cite{e5}, and BGE \cite{bge} have shown strong performance across multilingual and semantic retrieval benchmarks.

Despite recent progress, scientific source retrieval remains challenging due to domain mismatch between social media claims and formal scientific abstracts, multilingual semantic alignment difficulties, and limited training data for low-resource languages. Furthermore, dense retrieval alone may fail to capture fine-grained semantic interactions between claims and candidate documents. Neural reranking methods based on cross-encoders have therefore become increasingly important for improving retrieval precision by jointly modeling query-document interactions.

In this work, we propose a multilingual two-stage retrieval architecture combining fine-tuned dense retrieval with cross-encoder reranking for scientific source retrieval. Our system first retrieves semantically relevant candidate documents using multilingual dense embeddings and then applies neural reranking to improve top-ranked retrieval precision. We extensively evaluate multiple dense retrieval models, metadata configurations, retrieval depths, multilingual fine-tuning strategies, and reranking approaches.

Our experiments demonstrate that metadata enrichment, multilingual training, and fine-tuned cross-encoder reranking contribute to improved retrieval effectiveness across all languages. On the official development phase, our final system achieved MRR@5 scores of 0.6222 for English, 0.5112 for German, and 0.6254 for French with an overall average MRR@5 of approximately 0.59. On the official hidden evaluation phase, the system achieved an average MRR@5 score of 0.5565, demonstrating strong generalization capability across unseen multilingual scientific claims.

The main contributions of this work are summarized as follows:

\begin{itemize}
    \item We develop a multilingual dense retrieval pipeline for scientific source retrieval using fine-tuned transformer embedding models.
    
    \item We investigate the impact of metadata construction strategies, multilingual joint training, and German oversampling for improving retrieval quality.
    
    \item We evaluate the effectiveness of cross-encoder reranking for multilingual scientific claim retrieval.
    
    \item We provide extensive experimental analysis across development and hidden evaluation settings for the CLEF 2026 CheckThat! Task 1 benchmark.
\end{itemize}

\section{Related Work}

\subsection{Scientific Fact-Checking and Evidence Retrieval}

Scientific fact-checking has recently emerged as an important research area due to the rapid spread of misinformation and misleading scientific claims on social media platforms. Wadden et al.~\cite{scientificsurvey} provide a comprehensive survey of scientific fact-checking resources, datasets, and retrieval approaches. Their study highlights that scientific fact-checking differs substantially from traditional political fact-checking because scientific claims often require retrieval of long-form scholarly evidence, domain-specific reasoning, and evidence aggregation from multiple scientific publications. The survey further emphasizes the importance of semantic retrieval systems for identifying relevant scientific evidence from large-scale scholarly corpora.

Recent work has increasingly focused on evidence-based verification pipelines for health-related and scientific claims. Stammbach et al.~\cite{evidencefactchecking} proposed a retrieval-based framework for evidence-grounded fact-checking of health-related claims. Their system combines evidence retrieval with downstream verification components in order to assess claim reliability using scientific literature. The study demonstrates that retrieval quality plays a critical role in overall verification performance and that dense semantic retrieval models substantially outperform sparse lexical matching approaches when handling paraphrased or implicitly stated scientific claims.

The CLEF CheckThat! evaluation campaigns have also contributed significantly to research on multilingual fact-checking and scientific retrieval tasks. Schellhammer et al.~\cite{checkthat2026task1} introduced multilingual retrieval and verification tasks focusing on social media claims and evidence retrieval. Their work highlights the challenges posed by noisy user-generated content, multilingual variation, implicit scientific references, and semantically complex claims. The shared task setup further demonstrates the growing importance of multilingual semantic retrieval systems capable of linking social media claims to reliable scientific evidence.

Prior research has shown that dense neural retrieval architectures are particularly effective for semantic evidence retrieval tasks. Karpukhin et al.~\cite{dpr} introduced Dense Passage Retrieval (DPR), which demonstrated substantial improvements over traditional sparse retrieval methods by learning dense semantic representations for queries and documents. Similarly, Reimers and Gurevych~\cite{sbert} proposed Sentence-BERT, which enables efficient semantic similarity computation using transformer-based sentence embeddings. These approaches laid the foundation for modern dense retrieval systems widely adopted in fact-checking and evidence retrieval pipelines.

More recent retrieval architectures have focused on improving multilingual and contrastive representation learning for retrieval tasks. Izacard and Grave~\cite{contriever} proposed Contriever, an unsupervised dense retrieval model trained using contrastive learning objectives. Wang et al.~\cite{e5} introduced the multilingual-E5 family of embedding models using weakly supervised contrastive pretraining for multilingual semantic retrieval. Similarly, Xiao et al.~\cite{bge} proposed BGE embedding models optimized for high-quality semantic retrieval and ranking tasks. These multilingual embedding approaches have demonstrated strong performance across retrieval benchmarks involving semantic matching and multilingual generalization.

\subsection{Multilingual Scientific Retrieval}

Multilingual retrieval remains particularly challenging for scientific fact-checking tasks because semantic alignment must be preserved across languages while handling noisy social media text and specialized scientific terminology. Multilingual transformer models based on BERT~\cite{bert} and XLM-R~\cite{xlmr} have enabled significant advances in cross-lingual semantic representation learning. These models learn shared multilingual embedding spaces that facilitate semantic retrieval across multiple languages.

Despite recent advances, multilingual scientific retrieval still faces substantial challenges related to language imbalance, domain adaptation, and semantic ambiguity. Prior work has shown that retrieval performance often varies considerably across languages depending on data availability and linguistic complexity. Motivated by these findings, our work investigates multilingual dense retrieval and cross-encoder reranking strategies for scientific source retrieval across English, German, and French claims. Unlike prior work focusing primarily on verification, our system emphasizes multilingual scientific source retrieval through metadata enrichment, multilingual joint training, and reranking optimization.

\section{Methodology}

\subsection{Task Definition} 
The objective of the CLEF 2026 CheckThat! Task 1 is to retrieve the correct scientific publication corresponding to a social media claim. Given a query post and a shared collection of scientific publications, the system must rank the most relevant publication at the highest possible position.

Formally, given a social media claim $q$ and a collection of scientific publications
$D = \{d_1,d_2,\dots,d_n\}$,
the objective is to retrieve and rank the correct scientific publication $d^* \in D$ at the highest possible position.

The task is particularly challenging because social media claims are often short, noisy, paraphrased, and semantically incomplete. Furthermore, the multilingual nature of the dataset introduces additional complexity due to differences in linguistic structure across English, German, and French.

\subsection{System Overview}

\begin{figure}[htbp]
\centering
\includegraphics[width=\textwidth]{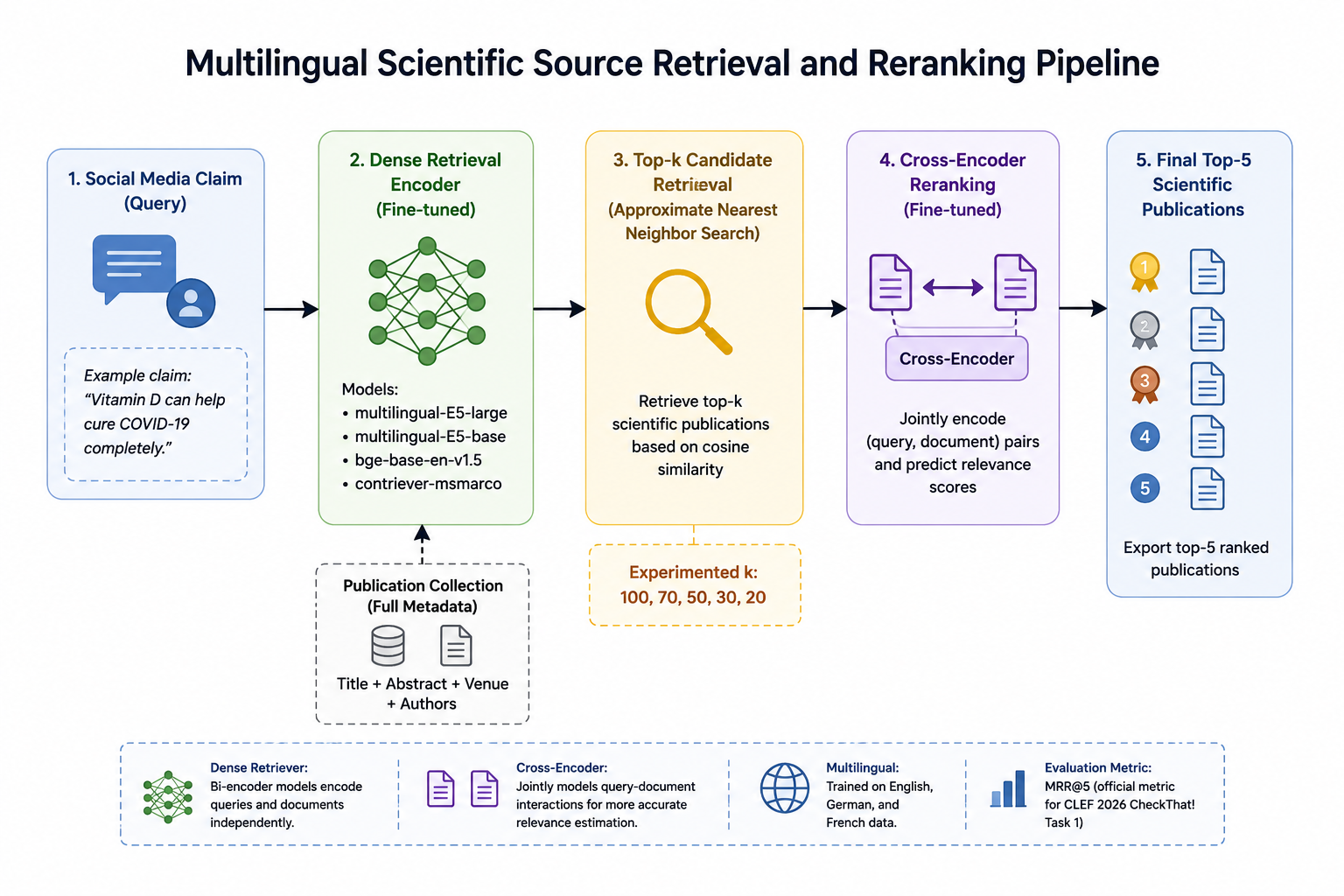}
\caption{Overview of the proposed multilingual retrieval and reranking pipeline.}
\label{fig:pipeline}
\end{figure}

Figure~\ref{fig:pipeline} illustrates the overall retrieval pipeline used in our system.

\FloatBarrier

The retrieval framework consists of the following stages:

\begin{enumerate}
    \item Query preprocessing
    \item Dense retrieval using fine-tuned bi-encoder models
    \item Top-k candidate retrieval using cosine similarity
    \item Cross-encoder reranking
    \item Final top-5 prediction generation
\end{enumerate}

In the first stage, dense retrieval models encode both social media claims and scientific publications into dense semantic vector embeddings within a shared embedding space. Given a query embedding $q$ and a document embedding $d$, semantic similarity is computed using cosine similarity:

\begin{equation}
\text{sim}(q,d)=\frac{q \cdot d}{\|q\| \, \|d\|}
\end{equation}

where $q \cdot d$ denotes the dot product between the embeddings and $\|\cdot\|$ represents the Euclidean norm. Candidate documents are ranked according to their cosine similarity scores with respect to the query embedding.

Unlike sparse lexical retrieval methods such as BM25~\cite{bm25}, dense retrieval enables semantic matching between paraphrased social media claims and formally written scientific abstracts~\cite{dpr,sbert,e5}. This is particularly important for scientific source retrieval where claims often omit technical terminology or reformulate scientific findings using simplified language. 

In the second stage, a cross-encoder reranking model jointly processes query-document pairs and refines the ranking of the retrieved candidates using fine-grained semantic interaction modeling.

Given a query-document pair $(q,d)$, the cross-encoder directly estimates a relevance score:

\[
s(q,d) = \mathrm{CrossEncoder}(q,d)
\]

where $s(q,d)$ represents the predicted semantic relevance between the social media claim and the candidate scientific publication. Candidate documents are then reranked according to these relevance scores.

\subsection{Dense Retrieval Models}

We experimented with multiple transformer-based dense retrieval architectures proposed in recent semantic retrieval literature, including Sentence-BERT~\cite{sbert}, Contriever~\cite{contriever}, multilingual-E5~\cite{e5}, BGE~\cite{bge}, and Dense Passage Retrieval (DPR)~\cite{dpr}.

The multilingual-e5 models demonstrated strong semantic retrieval capability across all three languages due to their multilingual embedding alignment and retrieval-oriented pretraining objectives. In contrast, English-focused models such as BGE and Contriever showed comparatively weaker multilingual generalization performance.

For E5-based models, the prefixes ``query:'' and ``passage:'' were applied for query and document encoding respectively.

Contriever models do not require query or passage prefixes and directly encode raw text inputs.

\subsection{Metadata Configurations}

We experimented with two different metadata configurations for scientific publication encoding.

\subsubsection{SIMPLE Configuration}

The SIMPLE configuration uses only the publication title for retrieval.

\subsubsection{FULL Configuration}

The FULL configuration concatenates multiple publication metadata fields including publication title, abstract, venue information, and author metadata.

The FULL configuration provides substantially richer semantic context compared to the SIMPLE title-only configuration. Incorporating abstracts, venues, and author information enables the retrieval model to capture broader semantic relationships between claims and scientific publications. Experimental results consistently demonstrated that the FULL configuration significantly improved retrieval effectiveness across all languages.

\subsection{Dense Retrieval Fine-Tuning}

Dense retrieval models were fine-tuned using the SentenceTransformers framework \cite{sbert} with contrastive learning objectives.

We used \texttt{MultipleNegativesRankingLoss} \cite{sbert}, which maximizes similarity between positive query-document pairs while minimizing similarity with non-matching pairs within the training batch.
Given a batch of positive query-document pairs $(q_i, d_i)$, the MultipleNegativesRankingLoss objective maximizes the similarity between matching pairs while minimizing similarity with other documents within the batch:

\[
\mathcal{L}
=
-\frac{1}{N}
\sum_{i=1}^{N}
\log
\frac{
\exp(\mathrm{sim}(q_i,d_i)/\tau)
}{
\sum_{j=1}^{N}
\exp(\mathrm{sim}(q_i,d_j)/\tau)
}
\]
where $\mathrm{sim}(q_i,d_j)$ denotes cosine similarity and $\tau$ is a temperature scaling parameter.

MultipleNegativesRankingLoss enables efficient contrastive learning by implicitly treating other documents within the same training batch as negative examples. This training strategy improves semantic discrimination capabilities within the embedding space without requiring explicit hard-negative mining.

During training, query and document embeddings were generated independently using transformer encoders. The bi-encoder retrieval architecture enables efficient large-scale retrieval since query and document embeddings can be computed independently and indexed offline prior to inference.

We fine-tuned both English-only and multilingual retrieval models. For multilingual training, English, German, and French datasets were combined into a joint training corpus.

To improve multilingual semantic alignment, multilingual fine-tuning was performed using combined English, German, and French training instances. Joint multilingual optimization encourages semantically similar claims and documents from different languages to occupy nearby regions within the embedding space, thereby improving multilingual retrieval robustness.

\subsection{Cross-Encoder Reranking}

Cross-encoder reranking is particularly beneficial for scientific retrieval tasks where multiple candidate publications may contain highly similar semantic content~\cite{crossencoder,dpr}. Unlike bi-encoder dense retrieval models that independently encode queries and documents, cross-encoders jointly process query-document pairs and capture deeper semantic interactions between social media claims and scientific publications. 
To improve retrieval precision, we therefore applied a cross-encoder reranking stage on top of the dense retrieval outputs. 
This allows more accurate ranking refinement at the cost of higher computational complexity.

The reranking pipeline operates as follows:

\begin{enumerate}
    \item Retrieve top-k candidate documents using dense retrieval
    \item Construct query-document pairs
    \item Compute cross-encoder relevance scores
    \item Sort candidate documents based on reranking scores
    \item Return final top-5 predictions
\end{enumerate}

We experimented with multiple retrieval depth values including top-100, top-70, top-50, top-30, and top-20 candidate documents.

Experimental results showed that retrieval depth had only a limited impact on reranking performance. Differences across retrieval depths were relatively small, with the top-50 configuration achieving the strongest development-set performance. Among the evaluated configurations, reranking the top-50 retrieved candidates provided the best development-set results and was therefore selected for the final system.

\subsection{Multilingual Joint Training}

To support multilingual retrieval, we jointly fine-tuned multilingual-e5-large \cite{e5} on English, German, and French datasets.

Initial multilingual experiments revealed that German retrieval performance consistently lagged behind English and French retrieval effectiveness. This performance gap may be attributed to increased linguistic complexity, domain-specific semantic variability, and comparatively weaker multilingual semantic alignment for German scientific claims. To address this imbalance, we oversampled German training examples during multilingual fine-tuning.

The oversampling strategy significantly improved German retrieval performance while preserving strong multilingual generalization across all supported languages. Oversampling increased the contribution of German training instances during optimization and improved language-specific representation learning. Importantly, the oversampling strategy improved German retrieval performance without substantially degrading English and French retrieval effectiveness, thereby preserving strong multilingual generalization capability.

\section{Experimental Setup}

\subsection{Dataset}

We used the dataset provided by the CLEF 2026 CheckThat! Task 1 organizers \cite{checkthat2026task1} for multilingual scientific source retrieval.\footnote{\url{https://huggingface.co/datasets/sschellhammer/CT26_Task1_SourceRetrievalForScientificWebClaims}}

The dataset consists of social media posts containing scientific claims in English, German, and French.

Each query post is associated with a single relevant scientific publication from a shared multilingual publication collection.

The scientific publication collection contains metadata fields including publication title, abstract, venue, authors, and publication identifiers.

We used the official training and development splits provided by the organizers for model training and evaluation.

\subsection{Evaluation Metric}

Retrieval effectiveness was evaluated using Mean Reciprocal Rank at cutoff 5 (MRR@5), which measures the ranking quality of the first relevant retrieved document among the top five predictions. MRR@5 is defined as:

\begin{equation}
\text{MRR@5} = \frac{1}{N}\sum_{i=1}^{N}\frac{1}{\text{rank}_i}
\end{equation}

where $\text{rank}_i$ denotes the rank position of the first correct retrieved document for query $i$, and $N$ represents the total number of evaluated queries.

MRR@5 is particularly suitable for scientific source retrieval tasks because it strongly rewards systems that rank the correct scientific publication at higher positions within the top retrieved candidates.

\subsection{Preprocessing}

Social media claims and publication metadata were preprocessed before retrieval training and inference.

For E5-based models, the prefixes ``query:'' and ``passage:'' were applied for query and document encoding respectively.

For Contriever models, raw text inputs were used directly without prefixes.

In the FULL metadata configuration, publication fields including title, abstract, venue, and authors were concatenated into a single retrieval document representation.

\subsection{Training Configuration}

Dense retrieval models were fine-tuned using the SentenceTransformers framework.

We used the following training configuration:

\begin{table}[htbp]
\centering
\caption{Dense Retrieval Training Configuration}
\label{tab:training_config}
\begin{tabular}{lc}
\toprule
Parameter & Value \\
\midrule
Loss Function & MultipleNegativesRankingLoss \\
Optimizer & AdamW \\
Learning Rate & 2e-5 \\
Batch Size & 32 \\
Epochs & 3 \\
Similarity Metric & Cosine Similarity \\
Max Sequence Length & 512 \\
\bottomrule
\end{tabular}
\end{table}

\FloatBarrier

Cross-encoder reranking models were fine-tuned using query-document relevance pairs generated from the retrieval dataset.

\subsection{Hyperparameter Tuning}

We experimented with multiple retrieval and reranking hyperparameters during training and inference.

For dense retrieval training, we evaluated different batch sizes, learning rates, and metadata configurations. For reranking, we experimented with varying retrieval depths including top-100, top-70, top-50, top-30, and top-20 candidate documents.

We additionally experimented with weighted score fusion between bi-encoder cosine similarity scores and cross-encoder relevance scores. However, standalone fine-tuned cross-encoder reranking produced the strongest retrieval effectiveness.

Our final system configuration was selected based on development phase MRR@5 performance.

\subsection{Hardware and Environment}

Experiments were conducted using NVIDIA T4 and NVIDIA P100 GPUs on Kaggle and Google Colab environments.

Dense retrieval training, multilingual fine-tuning, and reranking experiments were implemented using PyTorch, HuggingFace Transformers \cite{wolf2019huggingface}, SentenceTransformers \cite{sbert}, and NumPy.

\section{Results and Discussion} 

\subsection{Baseline Dense Retrieval Results}

We first evaluated several baseline dense retrieval architectures without task-specific fine-tuning. The objective of these experiments was to identify strong multilingual retrieval models for scientific source retrieval.

Table~\ref{tab:baseline_results} presents the baseline dense retrieval performance on the English development set before task-specific fine-tuning.

\begin{table}[htbp]
\centering
\caption{Baseline Dense Retrieval Results on the English Development Set}
\label{tab:baseline_results}
\begin{tabular}{lc}
\toprule
Model & MRR@5 \\
\midrule
multilingual-e5-large & 0.4986 \\
multilingual-e5-base & 0.4854 \\
bge-base-en-v1.5 & 0.4796 \\
contriever-msmarco & 0.4666 \\
\bottomrule
\end{tabular}
\end{table}


Among the evaluated baseline systems on the English development set, multilingual-e5-large achieved the strongest retrieval performance. The results indicate that multilingual transformer-based retrieval architectures provide strong semantic retrieval capabilities even before task-specific fine-tuning. 

\subsection{Fine-Tuned Dense Retrieval Results}

We next fine-tuned multiple dense retrieval architectures using the task training data. We additionally experimented with two metadata configurations:

\begin{itemize}
    \item SIMPLE: title-only retrieval
    \item FULL: title + abstract + venue + authors
\end{itemize}

Table~\ref{tab:finetuned_results} presents the fine-tuned dense retrieval performance on the English development set under SIMPLE and FULL metadata configurations.

\begin{table}[htbp]
\centering
\caption{Fine-Tuned Dense Retrieval Results on the English Development Set}
\label{tab:finetuned_results}
\begin{tabular}{lcc}
\toprule
Model & FULL & SIMPLE \\
\midrule
multilingual-e5-large & 0.5669 & 0.5596 \\
multilingual-e5-base & 0.5524 & 0.5451 \\
bge-base-en-v1.5 & 0.5842 & 0.5789 \\
contriever-msmarco & 0.5500 & 0.5413 \\
\bottomrule
\end{tabular}
\end{table}


Fine-tuning improved retrieval effectiveness across all evaluated retrieval models on the English development set. The FULL metadata configuration consistently outperformed the SIMPLE configuration, demonstrating the importance of metadata enrichment for scientific source retrieval.

Inclusion of publication abstracts, venue information, and author metadata provided a richer semantic context and improved retrieval quality. Based on the English development set results shown in Table~\ref{tab:finetuned_results}, the BGE-base-en-v1.5 model with the FULL metadata configuration achieved the strongest first-stage retrieval performance (MRR@5 = 0.5842) and was therefore selected as the base retriever for subsequent reranking experiments. The final multilingual system further incorporated cross-encoder reranking, multilingual joint training, and German oversampling, leading to the results reported in later sections.

\subsection{Model Variants}

We evaluated multiple dense retrieval architectures including multilingual-e5-large, multilingual-e5-base, bge-base-en-v1.5, and contriever-msmarco under SIMPLE and FULL metadata configurations.

In addition, we trained multilingual retrieval systems using joint multilingual fine-tuning across English, German, and French datasets. To improve the effectiveness of German retrieval, we applied German oversampling during multilingual training.

For reranking, we evaluated pretrained and fine-tuned cross-encoder architectures for ranking refinement.

\subsection{Impact of Cross-Encoder Reranking}

To further improve retrieval precision, we applied cross-encoder reranking on top of the retrieved candidate documents.

Table~\ref{tab:reranking_results} presents the reranking performance on the English development set.

\begin{table}[htbp]
\centering
\caption{Impact of Cross-Encoder Reranking on the English Development Set}
\label{tab:reranking_results}
\begin{tabular}{lc}
\toprule
System & MRR@5 \\
\midrule
Dense Retrieval Only & 0.5842 \\
Cross-Encoder Reranking & 0.5948 \\
Fine-Tuned Cross-Encoder & 0.6222 \\
\bottomrule
\end{tabular}
\end{table}


The results show that cross-encoder reranking provided a modest improvement over dense retrieval alone, increasing MRR@5 from 0.5842 to 0.5948. Fine-tuning the cross-encoder yielded a larger gain, achieving an MRR@5 of 0.6222. These findings highlight the benefit of modeling fine-grained semantic interactions between social media claims and candidate scientific publications.

Unlike bi-encoder retrieval models commonly used in dense retrieval systems \cite{dpr,sbert}, cross-encoders jointly process query-document pairs and directly model semantic interactions between claims and scientific publications \cite{crossencoder}.

\subsection{Retrieval Depth Analysis}

We investigated the impact of retrieval depth during reranking by varying the number of retrieved candidate documents.

Table~\ref{tab:topk_results} presents the reranking performance on the English development set for different retrieval depths.

\begin{table}[htbp]
\centering
\caption{Effect of Retrieval Depth During Reranking on the English Development Set}
\label{tab:topk_results}
\begin{tabular}{lc}
\toprule
Top-k Retrieval Depth & MRR@5 \\
\midrule
100 & 0.5931 \\
70 & 0.5934 \\
50 & 0.5948 \\
30 & 0.5941 \\
20 & 0.5942 \\
\bottomrule
\end{tabular}
\end{table}


The results indicate that retrieval depth had only a limited impact on reranking performance. Among the evaluated configurations, top-50 candidate retrieval achieved the highest MRR@5 score (0.5948), although the differences across retrieval depths were relatively small.

These findings suggest that retrieval depth was not a dominant factor in overall system performance for this task. Nevertheless, the top-50 configuration provided the strongest development-set performance and was therefore selected for the final system.

\subsection{Multilingual Joint Training and German Oversampling}

Initial multilingual training experiments showed that German retrieval performance significantly lagged behind English and French performance.

To address this imbalance, we oversampled German training examples during multilingual fine-tuning.

Table~\ref{tab:oversampling_results} presents the effect of German oversampling on multilingual development set retrieval performance.

\begin{table}[htbp]
\centering
\caption{Effect of German Oversampling During Multilingual Development Training}
\label{tab:oversampling_results}
\begin{tabular}{lcc}
\toprule
Language & Before Oversampling & After Oversampling \\
\midrule
German & 0.4758 & 0.5112 \\
French & 0.6154 & 0.6254 \\
\bottomrule
\end{tabular}
\end{table}

\FloatBarrier

German oversampling substantially improved German retrieval effectiveness while preserving strong multilingual retrieval performance across the remaining languages.

These results demonstrate that language-specific balancing strategies can effectively improve multilingual retrieval systems for lower-performing languages.

All reported experiments were conducted on the official development set provided by the shared task organizers.

\subsection{Development Phase Results}

We evaluated the final multilingual retrieval system on the official development phase datasets across English, German, and French.

Table~\ref{tab:multilingual_results} presents the multilingual retrieval performance.

\begin{table}[htbp]
\centering
\caption{Final Multilingual Development Phase Results}
\label{tab:multilingual_results}
\begin{tabular}{lc}
\toprule
Language & MRR@5 \\
\midrule
English & 0.6222 \\
German & 0.5112 \\
French & 0.6254 \\
\midrule
Average & 0.5862 \\
\bottomrule
\end{tabular}
\end{table}

\FloatBarrier

The multilingual retrieval system achieved strong retrieval performance across all supported languages. These results were obtained using the final two-stage retrieval pipeline consisting of the selected dense retriever, fine-tuned cross-encoder reranking, multilingual joint training, and German oversampling.

English and French achieved comparable retrieval effectiveness, while German retrieval performance remained comparatively lower. This may be due to linguistic variation, dataset imbalance, or differences in semantic representation quality across languages.

\subsection{Official Evaluation Phase Results}

Table~\ref{tab:eval_results} presents the final official evaluation phase results obtained on the hidden test set of the CLEF 2026 CheckThat! Task 1 competition.

Compared to the development phase results, the proposed multilingual retrieval and reranking pipeline demonstrated relatively stable generalization performance across unseen multilingual scientific claims. Although the hidden evaluation benchmark introduced additional semantic variability and more challenging retrieval scenarios, the overall reduction in retrieval effectiveness remained moderate.

English achieved the strongest evaluation performance with an MRR@5 score of 0.5961, followed by French with 0.5719 and German with 0.5014. The comparatively lower German performance remained consistent with observations from the development phase experiments. Nevertheless, multilingual joint training and German oversampling substantially improved German retrieval robustness.

Overall, the evaluation phase results confirm the effectiveness of combining fine-tuned multilingual dense retrieval with cross-encoder reranking for multilingual scientific source retrieval.

\begin{table}[htbp]
\centering
\caption{Official CLEF 2026 Evaluation Phase Results on the Hidden Test Set}
\label{tab:eval_results}
\begin{tabular}{lc}
\toprule
Language & MRR@5 \\
\midrule
English & 0.5961 \\
German & 0.5014 \\
French & 0.5719 \\
\midrule
Average & 0.5565 \\
\bottomrule
\end{tabular}
\end{table}


The multilingual retrieval system demonstrated relatively stable performance across both development and evaluation phases. English and French achieved stronger retrieval effectiveness, while German remained comparatively more challenging despite multilingual balancing and oversampling strategies. The average MRR@5 decreased from 0.5862 on the development set to 0.5565 on the hidden evaluation set, indicating reasonable generalization to unseen multilingual scientific claims.

\subsection{Comparison with Top Systems}

Table~\ref{tab:leaderboard_comparison} compares our final multilingual retrieval system with the top-performing systems on the official CLEF 2026 evaluation phase leaderboard.

\begin{table}[h]
\centering
\caption{Comparison with Top Systems on the Official CLEF 2026 Evaluation Phase}
\label{tab:leaderboard_comparison}
\begin{tabular}{lc}
\toprule
Team & Avg. MRR@5 \\
\midrule
claim2source & 0.7628 \\
team-outsider & 0.7580 \\
mattiaskvist & 0.7464 \\
CheckMate & 0.7194 \\
seupd2526-retrix & 0.7108 \\
\midrule
KhyontekAI & 0.5565 \\
\bottomrule
\end{tabular}
\end{table}

\FloatBarrier

While the highest-ranked systems achieved substantially stronger retrieval performance, our multilingual dense retrieval and reranking pipeline demonstrated stable multilingual generalization capability across English, German, and French scientific claims. The comparatively smaller performance gap between the development and hidden evaluation phases further indicates robust retrieval generalization without severe overfitting.

Several factors may have contributed to the performance gap relative to the top-ranked systems. Our approach relied primarily on dense retrieval and cross-encoder reranking, without incorporating hybrid sparse-dense retrieval, retrieval ensembles, hard-negative mining, or instruction-tuned retrieval models. In addition, the reranking stage used a relatively lightweight architecture compared to larger retrieval and ranking models commonly adopted in recent retrieval benchmarks. These limitations provide several directions for future work.

\section{Discussion}

Our experiments demonstrate that multilingual scientific source retrieval benefits from combining dense semantic retrieval with neural reranking architectures. While dense retrieval models effectively retrieve semantically relevant candidate publications \cite{dpr,sbert,e5}, cross-encoder reranking significantly improves top-ranked retrieval precision \cite{crossencoder} by modeling deeper token-level semantic interactions between claims and candidate scientific documents.

 The experimental results further highlight the importance of metadata construction for scientific retrieval tasks. Rich metadata representations containing abstracts, venue information, and author metadata consistently improved retrieval effectiveness, particularly for semantically paraphrased scientific claims \cite{scientificsurvey,evidencefactchecking} compared to title-only representations. In particular, publication abstracts provided important semantic context that enabled more robust matching between paraphrased social media claims and formal scientific descriptions. The results additionally suggest that retrieval quality depends not only on encoder representation quality but also on effective metadata construction and multilingual balancing strategies.

The retrieval depth experiments showed that performance remained relatively stable across different reranking depths. Although the top-50 configuration achieved the highest MRR@5 score, the observed differences were small. This suggests that retrieval depth had a limited influence on overall performance within the evaluated range, while still providing a practical basis for selecting the final reranking configuration.

The multilingual experiments revealed persistent performance differences across languages. English and French consistently achieved stronger retrieval effectiveness compared to German across both development and evaluation phases. The comparatively lower German performance may be attributed to increased linguistic complexity, compound word formation, and weaker semantic alignment within multilingual embedding spaces.

To mitigate this imbalance, we introduced German oversampling during multilingual joint fine-tuning. The oversampling strategy substantially improved German retrieval robustness while preserving stable multilingual performance across English and French retrieval settings. These findings demonstrate the effectiveness of language-balancing strategies for multilingual scientific retrieval tasks.

\subsection{Error Analysis}

Despite strong multilingual retrieval performance, several retrieval challenges remain unresolved.

One major source of retrieval error involves semantically similar scientific publications discussing closely related topics. In many cases, the dense retrieval system successfully retrieved highly relevant publications but failed to rank the exact ground-truth document at the top position. This issue was particularly common for scientific domains containing highly overlapping terminology and similar experimental findings.

We additionally observed that short or highly paraphrased social media claims often lacked sufficient contextual information for precise semantic matching. Claims containing incomplete scientific descriptions, informal wording, or implicit references to scientific concepts occasionally resulted in retrieval ambiguity.

Multilingual retrieval errors were more prominent for German claims. German compound word structures and domain-specific terminology sometimes reduced semantic alignment quality within the multilingual embedding space. Although German oversampling improved retrieval effectiveness, performance gaps between German and the other languages remained observable during both development and evaluation phases. Reranking occasionally amplified dense retrieval errors when the correct scientific publication was absent from the initial top-k candidate pool. This highlights the importance of strong first-stage retrieval recall for effective reranking performance.

The relatively stable performance between development and hidden evaluation phases indicates strong generalization capability of the proposed retrieval framework across unseen multilingual scientific claims. The overall MRR@5 decreased only from 0.5862 on the development phase to 0.5565 on the hidden evaluation phase, corresponding to a relatively small performance gap of 0.0297. This moderate reduction suggests that the proposed multilingual retrieval and reranking pipeline generalized effectively to unseen evaluation data without exhibiting severe overfitting behavior. Although the official hidden evaluation benchmark introduced additional semantic variability and more challenging retrieval scenarios, the performance reduction remained limited, further demonstrating the robustness of the proposed dense retrieval and reranking framework.

\section{Conclusion}

In this work, we presented a multilingual scientific source retrieval system for CLEF 2026 CheckThat! Task 1. Our approach combines fine-tuned dense retrieval models with cross-encoder reranking for retrieving scientific publications implicitly referenced in multilingual social media claims. We evaluated multiple retrieval architectures including multilingual-E5, BGE, and Contriever under different metadata configurations and multilingual training settings. Experimental results demonstrated that metadata enrichment, multilingual optimization, and fine-tuned reranking contributed to improved retrieval effectiveness across English, German, and French claims. The relatively small performance gap between the development phase (0.5862 MRR@5) and the hidden evaluation phase (0.5565 MRR@5) further indicates strong generalization capability and limited overfitting of the proposed retrieval pipeline.

In future work, we plan to investigate hybrid sparse-dense retrieval architectures, larger multilingual reranking models, hard negative mining strategies, instruction-tuned retrieval models, and ensemble retrieval systems for further improving multilingual scientific retrieval robustness and cross-lingual semantic alignment.



\section*{Acknowledgements}

We thank the CLEF 2026 CheckThat! organizers for providing the dataset and evaluation infrastructure.

\section*{Use of Generative AI}

During the preparation of this work, generative AI tools were used for language refinement, structural assistance, and drafting support. All experimental implementations, analyses, interpretations, and final scientific conclusions were verified and validated by the authors.

\bibliography{references}

\end{document}